\documentclass{webofc}
\usepackage[varg]{txfonts}   
\begin{document}
\title{Awkward Just-In-Time (JIT) Compilation:}
%
%
\subtitle{A Developer’s Experience}

\author{
    \firstname{Ianna}   \lastname{Osborne}\inst{1}\fnsep\thanks{\email{iosborne@princeton.edu}} 
    \and \firstname{Jim} \lastname{Pivarski}\inst{1}
    \and \firstname{Ioana} \lastname{Ifrim}\inst{1}
    \and \firstname{Angus} \lastname{Hollands}\inst{1}
    \and \firstname{Henry} \lastname{Schreiner}\inst{1}
}

\institute{Princeton University, Princeton, NJ 08544, USA
          }

\abstract{%
Awkward Array is a library for performing NumPy-like computations on nested, variable-sized data, enabling array-oriented programming on arbitrary data structures in Python. However, imperative (procedural) solutions can sometimes be easier to write or faster to run. Performant imperative programming requires compilation; JIT-compilation makes it convenient to compile in an interactive Python environment. Various functions in Awkward Arrays JIT-compile a user’s code into executable machine code. They use several different techniques, but reuse parts of each others’ implementations. We discuss the techniques used to achieve the Awkward Arrays acceleration with JIT-compilation, focusing on RDataFrame, cppyy, and Numba, particularly Numba on GPUs: conversions of Awkward Arrays to and from RDataFrame; standalone cppyy; passing Awkward Arrays to and from Python functions compiled by Numba; passing Awkward Arrays to Python functions compiled for GPUs by Numba; and header-only libraries for populating Awkward Arrays from C++ without any Python dependencies

}
\maketitle
\section{Introduction}
\label{intro}
Awkward Array library~\cite{Ref-AwkwardArray} offers its users an array-oriented programming style in a dynamically typed language. The advantage of the array-oriented calculations is that they fit the data analysis workflow better, considering the large amounts of data which a physicist typically loads into an interactive environment. Moreover,the code that performs one operation on all data is easy enough to write, then observe the intermediate results on distributions, and reiterate. The imperative code, on the other hand, is better for the stage after the interactive exploration, but the performant imperative programming requires compilation. The loops may take longer to write, especially when fitting into a Just-In-Time (JIT)-compiler and getting the compile-time types correctly, but are often self-explanatory and faster. It is not uncommon for a user analysis code to be expressed in a traditional high-performance, statically typed language - C++, and a statically typed language requires compilation.

The JIT-compilation makes it convenient to compile in an interactive Python environment. Several JIT-compilation techniques are used to achieve the desired acceleration. The Awkward Array functions JIT-compile a user’s code into executable machine code. They use different techniques, but reuse parts of each others’ implementations.

The techniques discussed in this article are focusing on integrating RDataFrame~\cite{Ref-RDataFrame}, cppyy~\cite{Ref-cppyy}, and Numba~\cite{Ref-Numba}, particularly Numba on GPUs. These include Awkward Arrays to and from RDataFrame conversions, the standalone cppyy integration, passing Awkward Arrays to and from Python functions compiled by Numba, passing Awkward Arrays to Python functions compiled for GPUs by Numba, and populating Awkward Arrays from C++ without any Python dependencies using a header-only library.

\section{Awkward Arrays and RDataFrame}
\label{sec-1}

The Awkward Arrays can be converted to the RDataFrame columns and the RDataFrame columns can be converted to Awkward Arrays as described in detail in~\cite{Ref-Awkward-RDataFrame} and~\cite{Ref-Awkward-RDataFrame-Tutorial}.

The user benefits from a faster execution of both the ROOT C++ functions and the user-defined pure C++ functions. Here is an example of such conversion shown beneath.

\begin{verbatim}
import awkward as ak

df = ak.to_rdataframe({"Events": array})

rdf = (
    df.Filter("Events.nMuon() == 2")
      .Filter("Events.Muon_charge()[0] != Events.Muon_charge()[1]")
      .Define("dimuon_mass", """
// this is C++
return std::sqrt(2 * Events.Muon_pt()[0] * Events.Muon_pt()[1]
    * (std::cosh(Events.Muon_eta()[0] - Events.Muon_eta()[1])
    - std::cos(Events.Muon_phi()[0] - Events.Muon_phi()[1])));
"""))
\end{verbatim}

The {\tt ak.to\_rdataframe} function presents a view of an Awkward Array as an RDataFrame source. The {\tt ak.from\_rdataframe} function converts the selected columns as native Awkward Arrays. The implementation leverages on a zero-copy Awkward Array view and implements all for-loops on data in C++.

The handle to this Array view is a lightweight 40-byte C++ object allocated on the stack. The generated RDataSource takes pointers into the original array data via this view. Next, the column readers are generated based on the run-time type of the views. Finally, the readers are passed to a generated source derived from {\tt ROOT::RDF::RDataSource}.

The {\tt ak.from\_rdataframe} function converts the selected columns as native Awkward Arrays. The templated C++ header-only implementation and the dynamically generated C++ code are used to extract the columns’ types and data.

\begin{verbatim}
import ROOT

df = ROOT.RDataFrame("Events",
    "root://eospublic.cern.ch//eos/opendata/cms/derived-data/"
    "AOD2NanoAODOutreachTool/Run2012BC_DoubleMuParked_Muons.root")

array = ak.from_rdataframe(
        df, columns=("Muon_charge", "Muon_eta", "Muon_mass",
                     "Muon_phi", "Muon_pt", "nMuon")
)
\end{verbatim}

\noindent This {\tt array} has type

\begin{verbatim}
type: 61540413 * {
    Muon_charge: var * int32,
    Muon_eta: var * float32,
    Muon_mass: var * float32,
    Muon_phi: var * float32,
    Muon_pt: var * float32,
    nMuon: uint32
}
\end{verbatim}

\noindent and some of its values look like

\begin{verbatim}
[{Muon_charge: [-1, -1], Muon_eta: [1.07, -0.564], ...},
 {Muon_charge: [1, -1], Muon_eta: [-0.428, 0.349], ...},
 {Muon_charge: [1], Muon_eta: [2.21], Muon_mass: [0.106], ...},
 ...,
 {Muon_charge: [-1, 1], Muon_eta: [-2.15, 0.291], ...}]
\end{verbatim}

On the one hand, the users who do their analysis entirely in Python ecosystem can benefit from faster execution using ROOT C++ functions, or pure C++, in an otherwise Awkward analysis at full speed. The performance cost of converting Awkward Arrays into RDataFrame is negligible, since it is a zero-copy view.

On the other hand, those who prefer C++, ROOT, and RDataFrame, have an ability to convert their data into Awkward Arrays in order to leverage the tools available in the wider Scientific Python ecosystem.

\section{Standalone cppyy}

Like ROOT and RDataFrame, {\tt cppyy} allows a user to write C++ and JIT-compile it to use it from Python. The C++ code can be included to Python from a C++ file or written directly as a Python string. But {\tt cppyy} can be installed without the entire ROOT package and Awkward Array's interface to it is more low-level: users can write C++ functions that operate on whole arrays, multiple arrays, and return arrays.

The {\tt ak.Array}, the Python class for all Awkward Arrays, implements a magic function {\tt \_\_cast\_cpp\_\_} that is called by {\tt cppyy} to determine a C++ type of the array. The Numba implementation~\cite{Ref-Awkward-Numba} is reused here to generate a C++ array view on demand. The generated ArrayView C++ class hashed type is registered as a {\tt cpp\_type} Python string attribute of the {\tt ak.Array} class. The {\tt cppyy} maps the C++ class type as a string to a Python type. The down side is that the user cannot redefine the function. Each function must have a unique name. This is similar to the PyROOT interpreter, because an earlier version of cppyy is used to bind Python and C++ in PyROOT.

The user does not need to know what {\tt cpp\_type} is - the {\tt cpp\_type} is generated on demand when the array needs to be passed to the C++ function.

\begin{verbatim}
array = ak.Array([
    [{"x": 1, "y": [1.1]}, {"x": 2, "y": [2.2, 0.2]}],
    [],
    [{"x": 3, "y": [3.0, 0.3, 3.3]}],
  ])

source_code_cpp = """
template<typename T>
double go_fast_cpp(T& awkward_array) {
 double out = 0.0;
 for (auto list : awkward_array) {
   for (auto record : list) {
     for (auto item : record.y()) {
       out += item;
     }
   }
 }
 return out;
}
"""
cppyy.cppdef(source_code_cpp)

out = cppyy.gbl.go_fast_cpp[array.cpp_type](array)
assert out == ak.sum(array["y"])
\end{verbatim}

The {\tt cppyy} version that is used must be 3.1 or later. The {\tt cppyy} library can construct an object of a previously declared type based on an arguments. This new feature is not available in the earlier versions. {\tt cppyy} implements an implicit instantiation from \texttt {\_\_cast\_cpp\_\_} returning a tuple.

\section{Awkward Arrays and Numba}
\label{sec-4}

Numba is a JIT-compiler of functions with Python syntax, and Awkward Arrays can be passed to and from Numba-compiled functions with a similar interface as {\tt cppyy}. Numba can be used in contexts in which acceleration is needed, but C++ is not---it allows users to write accelerated code in the same Python language as the rest of their code, albeit in a subset of that language (not all Python code can be compiled).


Numba infers the argument types at call time, and generates optimized code based on this information. Numba also compiles separate specializations depending on the input types.

\begin{verbatim}
import numba as nb, numpy as np

@nb.jit
def path_length(array):
  result = np.zeros(len(array), dtype=np.float32)
  for i, row in enumerate(array):
    result[i] = 0
    for j, val in enumerate(row):
      result[i] += val
  return result
\end{verbatim}

The implementation to pass Awkward Arrays to and from Python functions compiled by Numba defines a {\tt numba\_type} property of an {\tt ak.Array} that is a type of a generated C++ Array view, as discussed before. Awkward Arrays can be iterated over in Numba-compiled functions using zero-copy views.

\section{Python functions compiled for GPUs by Numba}

Numba can also compile code for Nvidia GPUs through LLVM's CUDA backend, and Awkward Arrays can be passed as arguments into such functions. Currently, the extension needs to be defined by a user in a function decorator (as {\tt extensions=[ak.numba.cuda]} below), but this requirement will be removed in a future version of Numba.

\begin{verbatim}
import cupy as cp

# make an Awkward Array of variable-length lists on the GPU with CuPy
N = 2**20
counts = cp.random.poisson(1.5, N).astype(np.int32)
content = cp.random.normal(0, 45, int(counts.sum())).astype(np.float32)
array = ak.unflatten(content, counts)

@nb.cuda.jit(extensions=[ak.numba.cuda])
def path_length(out, array):
    tid = nb.cuda.grid(1)
    if tid < len(array):
        out[tid] = 0
        for i, x in enumerate(array[tid]):
            out[tid] += x

blocksize = 256
numblocks = (N + blocksize - 1) // blocksize

result = cp.empty(len(array), dtype=np.float32)
path_length[numblocks, blocksize](result, array)
\end{verbatim}

\section{Header-only libraries}

To create Awkward Arrays in C++ without any dependence on Python, we provide a separate set of header-only libraries that implement an {\tt awkward::layoutbuilder}~\cite{RefH} that builds up an array by appending elements and then shares that array through basic C types (raw array buffers and a JSON-formatted string the convey structure).

\begin{verbatim}
#include "awkward/LayoutBuilder.h"
using awkward::layoutbuilder;

// instantiating layoutbuilder types (type aliases omitted for brevity)
enum Field : std::size_t {one, two};
using UserDefinedMap = std::map<std::size_t, std::string>;
UserDefinedMap fields_map({
    {Field::one, "one"},
    {Field::two, "two"}
});

RecordBuilder<
  RecordField<Field::one, NumpyBuilder<double>>,
  RecordField<Field::two, ListOffsetBuilder<int64_t,
      NumpyBuilder<int32_t>>>
> builder(fields_map);

// executable code to create and fill the layoutbuilder
auto& one_builder = builder.field<Field::one>();
auto& two_builder = builder.field<Field::two>();

one_builder.append(1.1);
auto& two_subbuilder = two_builder.begin_list();
two_subbuilder.append(1);
two_builder.end_list();

one_builder.append(2.2);
two_builder.begin_list();
two_subbuilder.append(1);
two_subbuilder.append(2);
two_builder.end_list();
\end{verbatim}

\section{Summary and future plans}

The Awkward Array library was designed around array-oriented programming, in part because imperative programming in Python is impractical at large scales. However, this library also provides many ways to access the same data---without copying it---in JIT-compiled contexts, enabling imperative computing at large scales.

While these bridges may be used directly by users, they are also pathways to future features. For example, automatic bindings from Kaitai Struct~\cite{Ref-kaitai} to Awkward Arrays are in development, and these rely heavily on the header-only {\tt awkward::layoutbuilder}. JIT-compilation in Julia~\cite{Ref-julia} is also in development, borrowing from the patterns set by C++ and Numba integration.

\section{Acknowledgment}
This work is supported by NSF cooperative agreement OAC-1836650 (IRIS-HEP) and NSF cooperative agreement PHY-2121686 (US-CMS LHC Ops).
%
%
%

\end{document}